\documentclass[12 pt, amsfonts,amsmath, amssymb,color]{article}

\usepackage{amsmath,amssymb,amsfonts,latexsym,float,graphics,epsfig}
\usepackage{subfig}
\usepackage{verbatim}
\usepackage{relsize}
\usepackage{hyperref}

\def\be{\begin{equation}}
\def\ee{\end{equation}}
\def\bea{\begin{eqnarray}}
\def\eea{\end{eqnarray}}

\def\R{\mathbb{R}}

\begin{document}

\renewcommand\theequation{\arabic{section}.\arabic{equation}}
\catcode`@=11 \@addtoreset{equation}{section}
\newtheorem{axiom}{Definition}[section]
\newtheorem{theorem}{Theorem}[section]
\newtheorem{axiom2}{Example}[section]
\newtheorem{lem}{Lemma}[section]
\newtheorem{prop}{Proposition}[section]
\newtheorem{cor}{Corollary}[section]

\newcommand{\ben}{\begin{equation*}}
\newcommand{\een}{\end{equation*}}
\title{\bf Contact geometry and thermodynamics of black holes in AdS spacetimes}
\author{
\bf Aritra Ghosh\footnote{E-mail: ag34@iitbbs.ac.in} \hspace{0.5mm} and Chandrasekhar Bhamidipati\footnote{E-mail: chandrasekhar@iitbbs.ac.in} \\
~~~~~\\
 School of Basic Sciences, Indian Institute of Technology Bhubaneswar,\\   Jatni, Khurda, Odisha, 752050, India\\
}

\date{ }

\maketitle

\begin{abstract}
In this paper we discuss a formulation of extended phase space thermodynamics of black holes in Anti de Sitter (AdS) spacetimes from the contact geometry point of view. Thermodynamics of black holes can be understood within the framework of contact geometry as flows of vector fields generated by Hamiltonian functions on equilibrium submanifolds in the extended phase space that naturally incorporates the structure of a contact manifold. Deformations induced by the contact vector fields are used to construct various maps among thermodynamic quantities. Thermodynamic variables and equations of state of Schwarzschild black holes are mapped to that of Reissner-Nordstr\"{o}m black holes in AdS, with charge as the deformation parameter. In addition, the equations of state of general black holes in AdS are shown to emerge from the high-temperature ideal gas limit equations via suitable deformations induced by contact vector fields. The Hamilton-Jacobi formalism analogous to mechanics is set up, and the corresponding characteristic curves of contact vector fields are explicitly obtained to model thermodynamic processes of black holes. Extension to thermodynamic cycles in this framework is also discussed.

\end{abstract}

\section{Introduction}

Formulation of the precise laws of black hole mechanics, in analogy with the laws of thermodynamics long ago~\cite{Bardeen:1973gs}, has paved the way for exciting proposals in general relativity~\cite{Bekenstein:1973ur}-\cite{Gibbons:1976ue}. For black holes in Anti de Sitter (AdS) spacetimes in particular, there has been a lot of activity, as they show interesting phase transitions, such as the Hawking-Page transition~\cite{Hawking:1982dh} and novel thermodynamic structure due to the presence of cosmological constant~\cite{Chamblin,Chamblin:1999hg, Caldarelli:1999xj}. The Hawking-Page phase transition between the unstable small black hole to stable large black hole phase is understood as confinement-deconfinement phase transition in the dual conformal field theory via AdS/CFT correspondence~\cite{Maldacena:1997re}-\cite{Witten:1998zw}. AdS/CFT correspondence remains one of the most important developments in the last few decades, in our efforts to unravel the connection between gravity and gauge theories in general.  Phase transitions of black holes provide a rich arena to explore these connections.\\

\noindent
More recently, with the revival of the proposal of treating the cosmological constant $\Lambda$, as being dynamical and as pressure, the first law of black hole mechanics has been modified by including new $PdV$ work terms, leading to exciting developments~\cite{Henneaux:1984ji}-\cite{Kubiznak:2016qmn}. With the presence of these pressure and volume terms, the enthalpy $H$ of the black hole takes the  centre stage, as opposed to internal energy $U$ in traditional treatments (where $\Lambda$ is a fixed parameter). Thus, the study of black holes in this novel approach where $\Lambda$ is treated as a thermodynamic variable, is considered to be {\em extended thermodynamic phase space} approach, and the holographic constructions in this setting go by the name of black hole chemistry.  Black holes and their thermodynamics in this extended phase space has seen tremendous activity in recent times (see~\cite{Kubiznak:2016qmn} for a review). Most importantly, the presence of new $PdV$ terms in the first law of black holes, allows for obtaining a suitable equation of state for each black hole with a study of PV critical behaviour, in analogy with standard thermodynamics~\cite{Kubiznak:2012wp}.
The objective of this paper is to set up a framework for understanding  black holes and their thermodynamics from a geometric point of view. The application of geometric methods to understand thermodynamics systems has been an interesting area of research with various approaches~ \cite{gibbs}-\cite{bravetti2015}. Analysis of thermodynamic geometry of black holes using Weinhold and Ruppeiner metrics~\cite{Weinhold,Ruppeiner:1995zz} and scalar curvatures is well known to have wide ranging applications to ideal gas, van der Waals fluids, quantum gases, Ising models etc.~\cite{Ruppeiner}-\cite{Quevedo:2006xk}, including using Gibbs approach to black hole thermodynamics~\cite{Cvetic:2018dqf}. More recently, with in the framework of extended black hole thermodynamics, these ideas have uncovered intriguing information about the underlying microscopic structure and phase transitions of black holes in AdS~\cite{Wei:2015iwa,Wei:2019uqg,Wei:2019yvs}. In this paper, we focus our attention on the contact structure, rather than the metric structure of the extended thermodynamic phase space, and uncover novel connections between various black holes in AdS.\\


\noindent
Contact geometry \cite{Geiges, Arnold} hasn't received much attention in physics literature until the recent years. This geometric setting is widely used to study thermodynamics \cite{RT1, RT3, RT2}, mechanical systems with Rayleigh dissipation \cite{CM1, CM2} as well as statistical mechanics \cite{SM}. Contact geometry is the odd dimensional counterpart of the more familiar symplectic geometry \cite{S}. The reason that symplectic geometry receives much greater attention in physics is because it forms the natural geometric embedding for conservative Hamiltonian dynamics which is often the starting point for the quantised theories. In contact geometry, the dynamics is described on an odd dimensional smooth manifold endowed with a specific structure in contrast to the symplectic case where the dynamics is described on an even dimensional symplectic manifold. The resulting dynamics in the contact scenario generalises that of the symplectic case and one obtains a set of contact Hamilton equations. In the mechanical case, these equations describe a dissipative system. Contact geometry forms a very natural setting for reversible thermodynamics. To see this, recall that in thermodynamics there are pairs of variables such as \(T\) and \(S\) which may be regarded as conjugate variables. Such pairs appear in this framework as coordinates in a thermodynamic phase space. However one should note an important difference with classical mechanics: in mechanics the phase space is always even dimensional whereas in the thermodynamic case the phase space turns out to be odd dimensional which can be seen from the expression below:
$$dU - TdS + PdV = 0 \, .$$
It is clear that \(U\) has no conjugate variable and hence the thermodynamic phase space in this case is five dimensional with local coordinates \(\{P,V,T,S,U\}\). Furthermore, in the case of black holes in extended phase space, enthalpy takes a centre stage and we work with the following relation:
$$dH - TdS - VdP=0 \, .$$
\noindent
The thermodynamic phase space assumes the structure of a contact manifold. In the equilibrium thermodynamics scenario, a particular thermodynamic system is restricted to an equilibrium submanifold in the thermodynamic phase space. Such a submanifold can be described the by the equation of state of the system. If \(f(V,T,....)=0\) be an equation of state for system then expressing it in terms of derivatives of the thermodynamic potential \(V = \partial H/\partial P\) and \(T = \partial H/\partial S\) we get:
$$
  f( \partial H/\partial P,\partial H/\partial S,....)=0 \, .
$$
This is known as the PDE of state \cite{Isodro} or a Hamilton-Jacobi equation \cite{Rajeev1}. Here the thermodynamic potential assumes the role of Hamilton's principal function.\\


\noindent
\textbf{Motivation and results:} The primary motivation of this paper is to formally develop a contact geometry formalism for discussing thermodynamics of black holes, in particular, in AdS. Although, geometric formulations of black hole thermodynamics have been explored before, as mentioned above, these studies are incomplete due to the lack of usual pressure $P$ and volume $V$ terms in the first law. With in the novel extended phase space thermodynamics and in the presence of traditional $PdV$ terms in black hole thermodynamics, the equations of state can be written in the form one is used to in standard thermodynamics. One of the motivations of this paper is to explore the possibility of representing various thermodynamic processes of black holes in the framework of contact Hamiltonian geometry. It is well known that Hamiltonian methods combined with canonical transformations of phase space variables is a powerful combination, helpful in solving nontrivial problems in mechanics, which are otherwise unsolvable. Thus, it is expected that a mechanical/geometrical approach to thermodynamics would be helpful in giving deeper insights into the phase space of thermodynamics variables and possibly provide novel relations among the systems themselves. Indeed, starting from the Schwarzschild black hole with relatively simple thermodynamic behaviour, we derive the thermodynamics of charged black holes in AdS as deformations induced by suitable contact vector fields in the thermodynamic phase space. We describe the high temperature ``ideal gas" limit of black holes in the contact framework and then provide a map to describe other black holes in AdS with non-trivial thermodynamic behaviour. Finally, we describe the generation of thermodynamic processes for black holes in the high temperature limit and extend this to discuss the Carnot cycle.\\

\noindent
The \underline{organization of the paper} is as follows. Section-(\ref{contactBH}) is mostly introductory where we collect the background required for rest of this paper on: contact geometry (in subsection-(\ref{contactG}) )and extended thermodynamics of black holes in AdS (in subsection-(\ref{ebh})). In subsection-(\ref{contactG}), we give a very brief outline of contact Hamiltonian systems and describe how they form a natural setting for reversible thermodynamics. We review the ideal gas and see how to achieve the van der Waals gas by considering deformations of the ideal gas by suitable contact vector fields. We then recall that a natural antisymmetric bracket structure between thermodynamic functions exists in the thermodynamic phase space. Subsection-(\ref{ebh}) contains recent results on the thermodynamics of black holes in AdS in extended phase space approach, where the cosmological constant is treated as dynamical variable, leading to new pressure and volume terms in the first law of black hole mechanics. Sections-(\ref{main}) and (\ref{HJeqn}) contain our main results and is devoted to studying the contact geometry approach to black hole thermodynamics. We describe the Schwarzschild black hole in the contact geometry setting and then deform it to achieve charged black holes in the AdS. We then consider the "high temperature limit" ideal gas behaviour of black holes and map it to Schwarzschild as well as charged AdS black holes, including the BTZ black hole in three dimensions. In section-(\ref{HJeqn}) we present the Hamilton-Jacobi equation for studying thermodynamics of black holes of arbitrary spacetime dimensions in the extended phase space and obtain characteristic curves representing various thermodynamic processes. This set up can be used to discuss the thermodynamic cycles in phase space, which form the natural setting to obtain a Carnot cycle. We end with remarks in section-(\ref{remarks}).

\section{Contact geometry and extended thermodynamics of black holes } \label{contactBH}

\subsection{Contact geometry} \label{contactG}
We shall now briefly digress on the basic aspects of contact geometry and hence, see how it forms the natural embedding for reversible thermodynamics. In contact geometry, the central concept is that of a contact manifold \((\mathcal{M},\eta)\) where \(\mathcal{M}\) is a smooth manifold of odd dimension \((2n+1)\) and \(\eta\) is a contact 1-form that satisfies the following condition of complete non-integrability:
\begin{equation}\label{nonintegrable}
  \eta \wedge (d\eta)^n \neq 0 \, ,
\end{equation}
where, \('\wedge'\) is the exterior product and \((d\eta)^n = d\eta \hspace{1mm} \wedge \hspace{1mm} d\eta \hspace{1mm} \wedge \hspace{1mm} ....... \hspace{1mm} \wedge \hspace{1mm} d\eta\) (n times). In this context, \(\eta \wedge (d\eta)^n\) can be identified as a standard volume form on \(\mathcal{M}\). The fact that \((\mathcal{M},\eta)\) satisfies the complete non-integrability condition [eqn (\ref{nonintegrable})] equivalently implies that one can write a Whitney sum decomposition of \(T\mathcal{M}\) in terms of the regular distributions \(ker(\eta)\) and \(ker(d\eta)\), i.e. one can write $T\mathcal{M} = ker(\eta) \oplus ker(d\eta)$.\\

\noindent
Associated with \(\eta\) there is a global vector field \(\xi\) known as the Reeb vector field determined uniquely by the following conditions:
\begin{equation}
  \eta(\xi) = 1, \hspace{5mm} d\eta(\xi,.) = 0 \, .
\end{equation}
The Reeb vector field dictates a natural splitting of the tangent bundle of \(\mathcal{M}\) given by:
\begin{equation}\label{splittingTM}
   T\mathcal{M} = L_{\xi} \oplus D \, ,
\end{equation}  where \(L_{\xi}\) is a vertical subspace generated by \(\xi\) and \(D\) is a horizontal distribution induced by the contact form:
\begin{equation}\label{D}
  D = ker(\eta) \, .
\end{equation}
If \((\mathcal{M},\eta)\) is a contact manifold and \(\Phi: \mathcal{M} \rightarrow \mathcal{M}\) be a diffeomorphism. Then \(\Phi\) is called a contact diffeomorphism if it follows that, \begin{equation} \Phi^*\eta=f\eta \, ,\end{equation} where \(f \in C^\infty(\mathcal{M})\) is some non-vanishing smooth function and \(\Phi^*\) is the pullback induced by \(\Phi\). The contact form is unique up to a scalar factor. Therefore, \(\eta\) and \(f\eta\) where \(f\) is some non-zero scalar function are equivalent in the sense that they lead to the same horizontal distribution [eqn (\ref{D})].\\

\noindent
It is always possible to find a set of local (Darboux) coordinates \((s,q^i,p_i)\) in the neighbourhood of any point on \(\mathcal{M}\) where \(i = 1,2, ...,n\). In these coordinates:
\begin{equation}
  \eta = ds - p_i dq^i \, .
\end{equation}
It can be shown that the following non-coordinate basis \cite{Splitting} is naturally adapted to the splitting of \(T\mathcal{M}\) [eqn (\ref{splittingTM})] :
\begin{equation}
  \{\xi,\hat{P}^i, \hat{Q}_i \}  = \bigg\{ \frac{\partial}{\partial s}, \frac{\partial}{\partial p_i}, p_i\frac{\partial}{\partial s} - \frac{\partial}{\partial q^i} \bigg\} \, .
\end{equation}
One can explicitly check that the operators \(\{\xi,\hat{P}^i, \hat{Q}_i \}\) obey the following commutation relations:
\begin{equation}
  [\hat{P}^i,\hat{Q}_j] = {\delta^i}_j \xi, \hspace{5mm} [\xi,\hat{P}^i] = 0, \hspace{5mm} [\xi,\hat{Q}_i] = 0 \, .
\end{equation}
This is the very well known algebra of the nth Heisenberg group, known as the Heisenberg algebra. We now define the dynamics on a the contact manifold. For every differentiable function \(h: \mathcal{M} \rightarrow \R\), there exists a so called contact vector field \(X_h\) generated by \(h\), and defined by the following relations:
\begin{equation}
 i_{X_h}\eta = -h; \hspace{3mm} i_{X_h}d\eta = dh - \xi(h)\eta \, .
\end{equation}
The generic contact vector field \(X_h\) in the Darboux coordinates takes the form:
\begin{equation}\label{contactfield}
  X_h = \bigg(p_i\frac{\partial h}{\partial p_i}-h\bigg)\frac{\partial}{\partial s} - \bigg(p_i \frac{\partial h}{\partial s}+\frac{\partial h}{\partial q^i}\bigg)\frac{\partial}{\partial p_i} + \bigg(\frac{\partial h}{\partial p_i}\bigg)\frac{\partial}{\partial q^i} \, .
\end{equation}
This means that the flow of \(X_h\) is given as:
\begin{equation}
  \dot{s} =  p_i\frac{\partial h}{\partial p_i} - h; \hspace{3mm} \dot{q}^i = \frac{\partial h}{\partial p_i}; \hspace{3mm} \dot{p}_i = -p_i \frac{\partial h}{\partial s} - \frac{\partial h}{\partial q^i} \, .
\end{equation}
These equations resemble the conservative Hamilton equations. Given a contact manifold \((\mathcal{M},\eta)\) and a differentiable function \(h:\mathcal{M} \rightarrow \R\), the triple \((\mathcal{M},\eta,h)\) shall be called a contact Hamiltonian system where \(h\) is the contact Hamiltonian function. For the usual mechanical case in one dimension, one can take \(h = H(q,p) + \alpha s\), where: \begin{equation}H(q,p) = \frac{p^2}{2m} + V(q) \, ,\end{equation} and \(\alpha\) being some constant. In that case the contact Hamilton equations for \(\dot{q}\) and \(\dot{p}\) give the usual mechanical equations with Rayleigh dissipation while the first equation gives:
\begin{equation}
  \dot{s} =  p \dot{q} - h \, ,
\end{equation}
which means in the case where \(s\) does not appear in \(h\), it is the Hamilton's principal function. The contact Hamiltonian \(h\) is not conserved along the flow of \(X_h\). It is straightforward to see from eqn (\ref{contactfield}) that:
\begin{equation}\label{doth}
 X_h(h)= \dot{h} = -h \frac{\partial h}{\partial s} \, .
\end{equation}
\noindent
On contact manifolds one can define a/ special class of submanifolds of maximal dimension whose tangent spaces are contained in the kernel of the contact form \(\eta\) at any point. More informally, they are solutions to the equation \(\eta=0\).\\

\noindent
Let \(L \subset \mathcal{M}\) be a submanifold of a contact manifold \((\mathcal{M},\eta)\) and \(\Phi: L \rightarrow \mathcal{M}\) be an inclusion map. If \(L\) be a maximal dimensional integral submanifold such that \(\Phi^*\eta=0\), then \(L\) is called a Legendre submanifold. It can be shown \cite{S} that the maximal dimension is \(n\) and locally the general form of such a submanifold \(L\) is given by:
\begin{equation}
   p_i = \frac{\partial F}{\partial q^i}, \hspace{3mm} q^j = -\frac{\partial F}{\partial p_j},  \hspace{3mm} s = F - p_j \frac{\partial F}{\partial p_j} \, ,
\end{equation} where \(I \cup J\) is a disjoint partition of the set of indices \(\{1,2,....,n\}, i \in I, j \in J\) and \(F=F(q^i,p_j)\) is a function of \(n\) variables known as the generator of Legendre submanifold \(L\). This implies that not all \(n\) dimensional submanifolds of \(\mathcal{M}\) are Legendre submanifolds: coordinates on a Legendre submanifold cannot include a conjugate pair.\\

\noindent
We remark that the Hamiltonian dynamics on a contact manifold is such that a Legendre submanifold \(L\) is invariant to the flow of \(X_h\) if and only if the contact Hamiltonian function \(h\) vanishes on \(L\)~\cite{RT1}. This means that if \(X_h\) be the contact vector field generated by \(h\), such that \(h|_L=0\) where \(L\) is a particular Legendre submanifold, and then, if the flow of \(X_h\) enters \(L\), it stays on \(L\). This follows from the fact that \(\dot{h}=0\) whenever \(h=0\) and hence \(X_h\) is tangent to the level surface for which \(h=0\). In other words, a contact vector field \(X_h\) is tangent to a Legendre submanifold \(L\) if and only if \(L \subset h^{-1}(0)\).

\subsubsection{Reversible thermodynamics}
We now show how contact Hamiltonian mechanics naturally describes classical thermodynamics. We define the thermodynamic phase space to be a contact manifold \((\mathcal{M},\eta)\). In terms of the Darboux coordinates, the contact 1-form can be expressed as:
\begin{equation}\label{contactstructure}
\eta = ds - p_Sdq^S - p_Vdq^V \, ,
\end{equation}
where \(S,V\) carry their usual meaning in thermodynamics. Recall that for a reversible thermodynamic process:
\begin{equation}\label{firstlaw}
 dU - TdS + PdV = 0 \, .
\end{equation}
Now that all thermodynamic systems at equilibrium satisfy eqn (\ref{firstlaw}), we then immediately identify them as Legendre submanifolds\footnote{We interchangeably use the terms Legendre submanifolds and equilibrium submanifolds.} of the contact thermodynamic phase space. This means that from eqn (\ref{contactstructure}) with \(q^S = S, q^V = V\) we get:
\begin{equation}
  s = U(S,V), \hspace{3mm} p_S = \frac{\partial U}{\partial S} = T, \hspace{3mm} p_V = \frac{\partial U}{\partial V} = -P \, . \hspace{3mm}
\end{equation}
Here the generator of the Legendre submanifold is the internal energy \(U = U(S,V)\). Referring to eqn (\ref{contactstructure}) it is clear that we could have chosen a different representation of the system where some other thermodynamic potential would have the role of the generating function for the Legendre submanifold. For example, a more relevant potential in the context of black hole thermodynamics (to be discussed in section-(\ref{ebh}) is the enthalpy \(H=H(S,P)\). Expressing eqn (\ref{firstlaw}) in terms of the enthalpy:
\begin{equation}
  dH - TdS -VdP = 0 \, ,
\end{equation}
and with \(q^S = S, q^V = P\) one immediately obtains:
\begin{equation}
  s = H(S,P), \hspace{3mm} p_S = \frac{\partial H}{\partial S} = T, \hspace{3mm} p_V = \frac{\partial H}{\partial P} = V \, .
\end{equation}
These thermodynamic systems can interpreted geometrically~\cite{RT1,RT2} as the triplet \((\mathcal{M},\eta,L)\) where \(L\) is the Legendre submanifold corresponding to that particular system in the thermodynamic phase space \((\mathcal{M},\eta)\). A contact vector field \(X_h\) generated by a contact Hamiltonian \(h\) can be considered as a generator of a thermodynamic process on a Legendre submanifold \(L\), if the Hamiltonian function vanishes on the appropriate submanifold, i.e. \(h|_L=0\). The flow of such a vector field is tangent to the equilibrium submanifold and the flow stays on \(L\). Therefore, a thermodynamic system undergoing a particular transformation is interpreted as the quadruple \((\mathcal{M},\eta,h,L)\) where \(h=0\) on \(L\). For discussions on thermodynamic processes for black holes in section (\ref{TProcesses}), we recall two important examples below.
\subsubsection{The classical ideal gas}
The relevant thermodynamic variables for describing the ideal gas are given as \(\{U,T,S,P,V,\mu,N\}\), in a seven dimensional thermodynamic phase space. The following relation:
\begin{equation}\label{firstlawideal}
  dU - TdS + PdV - \mu dN = 0 \, ,
\end{equation}
is identified with the vanishing of a contact form defining an equilibrium submanifold. Therefore, the conjugate variable pairs are: \((q^1,p_1) \rightarrow (S,T)\); \((q^2,p_2) \rightarrow (V,-P)\); \((q^3,p_3) \rightarrow (\mu,N)\) with \(s=U\). The internal energy \(U\) is the generating function for the appropriate Legendre submanifold which may be expressed as:
\begin{equation}\label{internal energy}
  U(S,V) = U_0 \exp[S/CNR]V^{-1/C}N^{1+1/C} \, ,
\end{equation}
where, \(C\) is the specific heat at constant volume and \(U_0 > 0\) is an appropriate constant. We can easily recover the ideal gas equation \(PV=NRT\) from eqn (\ref{internal energy}) by taking first derivatives. Consider the contact Hamiltonian \(h = TS - RNT +\mu N - U\). Then the contact Hamiltonian vector field [eqn (\ref{contactfield})] takes the form:
\begin{equation}\label{ideal}
  X_h = -U\frac{\partial}{\partial U} + (RN-S)\frac{\partial}{\partial S} - P\frac{\partial}{\partial P} - N\frac{\partial}{\partial N} - RT\frac{\partial}{\partial \mu} \, .
\end{equation}
This implies an isochoric-isothermal transformation since \(\dot{T}=\dot{V}=0\) while other thermodynamic variables evolve:
\begin{equation}
  \dot{U}=-U, \hspace{2mm} \dot{S}= NR-S, \hspace{2mm} \dot{P}=-P, \hspace{2mm} \dot{N}=-N, \hspace{2mm} \dot{\mu}=-RT \, .
\end{equation}
Therefore, the evolution of the thermodynamic variables is:
\begin{equation}\label{idealisochoricisothermal}
  U(\tau)=U_0e^{-\tau}, \hspace{2mm} S(\tau)= (RN_0\tau - S_0)e^\tau, \hspace{2mm} P(\tau)=P_0e^{-\tau}, \hspace{2mm} N(\tau)=N_0e^{-\tau}, \hspace{2mm} \mu(\tau)=\mu_0 -RT_0\tau \, ,
\end{equation}
where \(\tau \in \R\). One should note that for the ideal gas, \(PV = NRT\) and hence \(h=TS-PV + \mu N - U =0\). Therefore, the Legendre submanifold \(L\) representing the gas is invariant under the flow of \(X_h\), i.e. \(X_h\) is tangent to \(L\). The vector field \(X_h\) [eqn (\ref{ideal})] therefore represents an isochoric-isothermal transformation. It is straightforward to check that the evolution equations [eqns (\ref{idealisochoricisothermal})] satisfy the ideal gas equation of state and the equipartition theorem \(U = 3NK_BT/2\).

\subsubsection{The van der Waals gas}
In the previous example \(X_h\) was tangent to the Legendre submanifold \(L\) representing the system and hence could be regarded as the generator of the thermodynamic process. However there can be cases where the contact Hamiltonian vector field is not tangent to the Legendre submanifold, i.e. the Legendre submanifold is not invariant. Then \(X_h\) cannot be treated as a generator of a thermodynamic process but rather as a generator of a family of thermodynamic systems \cite{therm}. We now consider the transformations that map the ideal gas into a real gas. First consider \(h_1 = a/V\) where \(a>0\). The contact Hamiltonian vector field is therefore: \(X_{h_1} = -(a/V)\partial/\partial U - (a/V^2)\partial/\partial P\). Thus internal energy and pressure are the only two non-conserved thermodynamic variables which evolve as:\begin{equation}U = U_0 - \frac{a}{V_0}\tau_1, \hspace{3mm} P = P_0 - \frac{a}{V_0^2}\tau_1; \hspace{5mm} \tau_1 \in \R \end{equation}\label{idealdeformpu} One concludes that \(X_{h_1}\) maps the ideal gas into an interacting gas of point particles. Second, consider \(h_2 = -bP\) where \(b>0\). Then the contact Hamiltonian vector field becomes: \(X_{h_2} = b\partial/\partial V\). This means that except for the volume that evolves linearly, all other thermodynamic variables are constant. The volume evolves as: \begin{equation}V = V_0 + b\tau_2; \hspace{5mm} \tau_2 \in \R\end{equation} Therefore one may say that \(X_{h_2}\) maps the ideal gas into a non-interacting gas with finite molecular volume. The van der Waals gas can be obtained if we consider the two transformations \(X_{h_2}\) and \(X_{h_1}\) successively. In this case, the equation of state takes the form:
\begin{equation}
  \bigg(P + \frac{a}{V^2}\tau_1\bigg)(V-b\tau_2)=NRT \, .
\end{equation}
This is a two-parameter equation of state induced by the action of \(X_{h_2}\) followed by \(X_{h_1}\). There is in addition a deformation induced in \(U\) by the vector field \(X_{h_1}\). Therefore, in terms of the ideal gas internal energy \(U_0=3Nk_BT/2\) we have the internal energy of the van der Waals gas to be:
\begin{equation}\label{vanderwaalsU}
  U = \frac{3Nk_BT_0}{2} - \frac{a}{V_0}\tau_1 \, .
\end{equation}
This is in fact the equipartition theorem for the van der Waals gases, which is generated as a deformation on the ideal gas. The deforming vector field can thus be viewed as a generator of a family of thermodynamic systems, i.e. a family of Legendre submanifolds. As a matter of fact, the transformations \(X_{h_1}\) and \(X_{h_2}\) don't commute which can be checked from their non-vanishing commutator bracket. Therefore the transformation \(X_{h_1}\) followed by \(X_{h_2}\) shall lead to a different equation of state which would altogether differ from that of \(X_{h_1 + h_2}\) (see \cite{therm} for details on these examples although their conventions are slightly different from ours).

\subsection{Lagrange brackets} \label{Lagrange}
A contact manifold \((\mathcal{M},\eta)\) has naturally associated with it, a Jacobi structure \cite{Jacobi,Jacobi2} which can be defined by a bilinear map, \(\{.,.\}:C^\infty(\mathcal{M}) \times C^\infty(\mathcal{M}) \rightarrow C^\infty(\mathcal{M})\) known as a Jacobi bracket that is antisymmetric and as a consequence obeys the Jacobi identity. The Jacobi bracket generically is of the form:
\begin{equation}\label{jacobi}
  \{h_1,h_2\} = \Lambda(dh_1,dh_2) + h_1E(h_2) - h_2E(h_1) \, ,
\end{equation}
where, \(E\) is a vector field and \(\Lambda\) is a 2-contravariant bivector field on \(\mathcal{M}\) such that the following relations hold:
\begin{equation} [\Lambda,\Lambda] = 2E \wedge \Lambda; \hspace{3mm} [\Lambda,E]=0 \end{equation} \, , where \([.,.]\) is the Schouten-Nijenhuis bracket \cite{Jacobi3}.
In the contact geometry case, \(E\) is identified with the Reeb vector field \(\xi\) and then putting \(\Lambda(dh_1,dh_2)=d\eta(X_{h_1},X_{h_2})\) we get from eqn (\ref{jacobi}) a Jacobi bracket which in local coordinates is expressed as:
\begin{equation}
\{h_1,h_2\}=  h_1 \frac{\partial h_2}{\partial s} - \frac{\partial h_1}{\partial s} h_2  + p_i\bigg(\frac{\partial h_1}{\partial s} \frac{\partial h_2}{\partial p_i} - \frac{\partial h_1}{\partial p_i} \frac{\partial h_2}{\partial s}\bigg) + \frac{\partial h_1}{\partial q^i} \frac{\partial h_2}{\partial p_i} -\frac{\partial h_1}{\partial p_i}\frac{\partial h_2}{\partial q^i}.
\end{equation}
This bracket is known as the Lagrange bracket \cite{Rajeev1, Rajeev2}. It does not satisfy the Leibniz rule and is hence not a Leibniz bracket. As a consequence, the Lagrange bracket of constant with an arbitrary function may not vanish:
\begin{equation}
  \{1,F\}=\xi(F) \, .
\end{equation}
This allows one to define Lagrange bracket relations between pairs of thermodynamic variables \(q^S = S, q^V = P, p_S = T, p_V = V\) for \(s=H(S,P)\):
\begin{equation}
  \{S,T\}=1; \hspace{3mm}  \{P,V\}=1 \, .
\end{equation}
Lagrange brackets provide a nice starting point for discussing quantum thermodynamics and thermodynamic uncertainty relations~\cite{Rajeev2}.

\subsection{Black hole thermodynamics in extended phase space} \label{ebh}
We now summarise recent developments in thermodynamics of black holes in the novel extended phase space approach, where the cosmological constant is considered to be dynamical, giving pressure $P$. Let us start from the Bekenstein-Hawking formula for entropy of black holes given is~\cite{Bardeen:1973gs}-\cite{Gibbons:1976ue}:
\begin{equation}
S_{BH} = \frac{A}{4}.
\end{equation}
Including the quantum mechanical considerations, Hawking proved that black holes do emit thermal radiation (\textit{Hawking radiation}) at  temperature\footnote{We  use geometrical units where $G= c=\hbar = k_B = 1.$ }:
\begin{equation}
T_H = \frac{\kappa}{2\pi}.
\end{equation}
The thermodynamic quantities, \textit{Hawking temperature} $T_H$ and black hole entropy $S_{BH}$, suggest that the relationship between the laws of black holes and laws of thermodynamics is much  more than  an analogy.
This becomes evident when the \textit{first law of thermodynamics} for semi-classical black holes is written as:
\begin{equation}
\label{first law non extended}
dM = TdS + \Phi dQ,
\end{equation}
where, mass $M$  is identified with internal energy  $U$ of the black hole. It is notable that there is no $PdV$ term in the  first law  (equation-(\ref{first law non extended})), which is a familiar term in ordinary thermodynamics. This issue is resolved in the extended thermodynamic description of black holes, where the cosmological constant is considered dynamical~\cite{Henneaux:1984ji,Teitelboim:1985dp,Henneaux:1989zc,Caldarelli:1999xj,Wang:2006eb,Sekiwa:2006qj,LarranagaRubio:2007ut,Padmanabhan:2002sha}. Pressure is defined as:
\begin{equation}  \label{PL}
P = -\Lambda/8\pi \, ,
\end{equation}
where \(\Lambda\) is the dynamical cosmological constant. This is now incorporated in several works~\cite{Kastor:2009wy}-\cite{Kubiznak:2016qmn}. An important justification for this new term came from the work of Kastor, Ray and Traschen~\cite{Kastor:2009wy} who showed using scaling arguments, that the Smarr~\cite{Smarr:1972kt} relations necessitate such a term. In Anti de Sitter spacetime, the mass of a black hole is more appropriately interpreted as enthalpy $H$, rather than traditional identification  as internal energy $U$\cite{Kastor:2009wy}:
 \begin{equation}
 M = H (S,P) = U(S,V) + PV \, .
 \end{equation}
With the presence of pressure and thermodynamic volume above, the first law of black hole thermodynamics takes the standard form,
\begin{equation} \label{dU}
{d U = T d S +\Phi d Q - PdV} \, ,
\end{equation}
which can be obtained from the Legendre transform of
\begin{equation}\label{dHM}
d M = dH = T d S +\Phi d Q + V dP \, ,
\end{equation}
where, \(Q\) is the electric charge\footnote{We shall work in the fixed \(Q\) ensemble and hence put \(dQ=0\) in subsequent discussions.} of the black hole, \(\Phi\) is the electric potential and all other symbols having their usual meaning in thermodynamics. Equations such as equation-(\ref{dHM}) involving pressure and volume terms, were discussed before in different contexts~\cite{Caldarelli:1999xj,Wang:2006eb,Sekiwa:2006qj,LarranagaRubio:2007ut,Padmanabhan:2002sha}. \\

\noindent
To understand the concept of thermodynamic volume $V$, it is useful to work it out for a sample case of a non-rotating (static) black hole in four dimensions, namely, the Schwarzschild black hole in AdS. The corresponding line element is:
\begin{equation}
ds^2 = -f(r) dt^2 + f^{-1}(r)dr^2 + r^2 d\Omega^2,
\end{equation}
with
\begin{equation} \label{fdef}
f(r) = 1 -\frac{2 m}{r} + \frac{r^2}{l^2} ,
\end{equation}
where $\Lambda = -3/l^2$ and $d\Omega^2=d\theta^2+\sin^2\theta d\phi^2$. The location of horizons $r_+$, can be found from the zeros of the lapse function $f(r)$. The known relations are:
\begin{equation}\label{SandP}
S={\pi r_+^2},  \qquad m= \frac {r_+}{2} \left(1  + \frac{r_+^2}{l^2}  \right),
\end{equation}
where the Arnowitt-Deser-Misner (ADM) mass is $M=m$. Together with the new expression for pressure in eqn. (\ref{PL}), the enthalpy, $H(S,P)$ can now be obtained from equation-\eqref{SandP} to be~\cite{Kastor:2009wy}:
\begin{equation} \label{HSCH}
H(S,P)=M=m=\frac {1} {2} \left(\frac {S}{\pi}\right)^{\frac 1 2}
\left(1+\frac {8 S P} {3} \right).
\end{equation}
The temperature and volume can also be obtained using standard thermodynamic relations as~\cite{Dolan:2010ha,Dolan:2011xt,Dolan:2012jh}:
\begin{eqnarray} \label{TH}
T&=&\left(\frac{\partial H}{\partial S} \right)_P =\frac {1} {4}\left(\frac{1}{\pi S }\right)^{\frac 1 2}
(1+8   P S)=\frac{ (1 +  \frac{3  r_+^2}{l^2})}
{4\pi r_+}\;, \\
\label{VH}
V&=& \left(\frac{\partial H}{\partial P} \right)_S  \frac 4 3 \frac{S^{\frac 3 2}}{\sqrt{\pi}} = \frac {4 \pi r_+^3}{3}.
\end{eqnarray}
It can be checked that the above equations for temperature and volume are same as the ones in standard black hole thermodynamic treatments. In particular, volume $V$ turns out to be similar to the putative geometric volume and is not independent from entropy\footnote{However, this is just a coincidence, specifically valid for static black holes. In general, for example for rotating black holes, entropy $S$ and volume $V$ are not straightforwardly related to each other~\cite{Dolan:2010ha,Dolan:2011xt,Dolan:2012jh}.}.
 The relevant thermodynamic variables satisfy the Smarr relation:
\begin{equation}\label{Smarr}
  (d-3)M  = (d-2)TS + (d-3)Q\Phi - 2PV \, ,
\end{equation} where \(d\) is the number of spacetime dimensions. As we will see in later sections, in the contact geometry framework, $S$ and $V$ may be treated as independent, to start with.
The above analysis becomes more interesting for charged black holes in AdS, as this system gets mapped exactly to the day to day van der Waals fluid system in usual thermodynamics~\cite{Chamblin,Chamblin:1999hg,Kubiznak:2012wp}.
The enthalpy, in terms of the thermodynamic volume $V$ and pressure $P$ for charged black holes are known to be~\cite{Dolan:2010ha,Dolan:2011xt,Dolan:2012jh}:
\begin{equation}\label{HRN}
H(S,P)=M=m=\frac {1} {6\sqrt{\pi}} S^{-\frac{1}{2}}
\left(8 P S^2 + 3S+3\pi Q^2 \right) \, .
\end{equation}
The temperature of the Black Hole is:
\begin{equation} \label{TRN}
T =  \frac{1}{4\sqrt{\pi}} \, S^{-3/2} \left( 8 P S^2 + S - \pi Q^2 \right)\, .
\end{equation}
These quantities given above satisfy the first law of black hole thermodynamics in extended phase space (including $P$ and $V$). Motivated by this, in a remarkable work, Kubiznak and Mann studied the full $P-v$ critical behavior of charged black holes in AdS in a fixed charge ensemble in $(P,T)$-plane~\cite{Kubiznak:2012wp}, starting from an equation of state:
\begin{equation} \label{eos1}
P=\frac{T}{v}-\frac{1}{2\pi v^2}+\frac{2Q^2}{\pi v^4} \, .
\end{equation}
Here,
\begin{equation}
v=2r_+\, =2\left(\frac{3{V}}{4\pi}\right)^{1/3} \, ,
\end{equation}
is the specific volume\footnote{We set the Planck length, \(l_P = 1\).}, considered to be associated to the van der Waals fluid (rather than the thermodynamic volume $V$). Notably, the behaviour is similar to the van der Waals gas in ordinary thermodynamics with exact mapping of critical exponents as well~\cite{Kubiznak:2012wp}. The equilibrium condition \(\eta = dH - TdS - VdP - \Phi dQ=0\) can be identified with a contact structure vanishing on an equilibrium subspace. It should be remarked that eqns (\ref{Smarr}) and (\ref{dHM}) are related by a scaling argument. The specific volume \(v\) is related to the horizon radius \(r_+\) as:\begin{equation}\label{specificvolume} 4r_+ = (d-2)v \, . \end{equation}
It is clear that the conjugate pairs of variables are \(V\) and \(P\), \(T\) and \(S\), \(\Phi\) and \(Q\). Hence, the following Lagrange bracket relations between these pairs hold:
\begin{equation}\label{Lagrange}
  \{S,T\}=1; \hspace{3mm} \{P,V\}=1; \hspace{3mm} \{Q,\Phi\}=1 \, .
\end{equation}

\section{Contact geometry description of black holes} \label{main}

\subsection{Contact vector fields and associated flows for black holes}
We now describe the thermodynamics of black holes in the AdS spacetime in the contact geometry framework. In particular, we show that one can construct a Hamiltonian description of flows in the thermodynamic phase space that are consistent with the thermodynamic equations of state. Starting from the Schwarzschild black hole with zero electric charge, we see that the equation of state of charged black holes, containing the charge term can be obtained by considering suitable deformations induced by contact Hamiltonian vector fields. Hence, we shall describe the ideal gas limit of black holes and deform it to obtain the thermodynamic equations of some of the AdS black holes.

\subsubsection{Schwarzschild black holes in AdS}\label{SBH}
As discussed in last subsection, the Enthalpy and Temperature of the Schwarzschild black hole with zero electric charge (\(Q=0\)) are:
\begin{eqnarray}\label{Schwarzchild}
  H(S,P) &=& \frac{1}{6\sqrt{\pi S}}(8PS^2 + 3S) \, , \nonumber \\
  T  &=& \frac{1}{4 \sqrt{\pi}}S^{-3/2} \left( 8 P S^2 + S \right).
\end{eqnarray}
The thermodynamic equation of state can be easily calculated by taking first derivatives of the enthalpy. In terms of the specific volume \(v\) one obtains the equation of state:
 \begin{equation}\label{Schwarzchildeqn}
  P = \frac{T}{v} - \frac{1}{2\pi v^2} \, ,
\end{equation}
where in \(d=4\), we have \(v=2r_+\).
We consider a contact Hamiltonian function \(h = 2(TS - PV) - H\). Then the generic contact Hamiltonian vector field \(X_h\) takes the form:
\begin{equation}\label{vectorfieldschwarzchild}
  X_h = H\frac{\partial}{\partial H} - T\frac{\partial}{\partial T} + 3V\frac{\partial}{\partial V} + 2S\frac{\partial}{\partial S} - 2P\frac{\partial}{\partial P} \, ,
\end{equation}
so that the flow of \(X_h\) is given by:
\begin{equation}\label{flowschwarzschild}
  \dot{H}=H, \hspace{3mm} \dot{T}=-T, \hspace{3mm} \dot{V}=3V, \hspace{3mm} \dot{S}=2S, \hspace{3mm} \dot{P}=-2P \, .
\end{equation}
In particular \((P,V,T)\) of the black hole evolve as:
\begin{equation}\label{schwarzschildcharacteristics}
  P(\tau) = P_0e^{-2\tau}, \hspace{3mm} V(\tau) = V_0e^{3\tau}, \hspace{3mm} T(\tau) = T_0e^{-\tau} \, ,
\end{equation}where \(\tau \in \R\).
In \(d=4\) one may write for the specific volume: $ v(\tau) = v_0 e^\tau $ with \(v_0 = 2(3V_0/4\pi)^{1/3}\). Then it can be easily verified that these flow equations satisfy the equation of state [eqn (\ref{Schwarzchildeqn})] for the Schwarzschild black hole.\\

\noindent
From the Smarr relation [eqn (\ref{Smarr})] in the \(d=4\) case we find that, \(h = 2(TS - PV) - H = 0\) and therefore, the Legendre submanifold representing the black hole is invariant to the flow of \(X_h\). In other words, the contact vector field \(X_h\) is tangent to the corresponding Legendre submanifold. The enthalpy [eqn (\ref{Schwarzchild})] and the Hamiltonian \(h = 2(TS - PV) - H\) completely defines the thermodynamics of a Schwarzschild black hole within the setting of contact geometry.

\subsubsection{BTZ black holes in AdS}
The Banados-Teitelboim-Zanelli (BTZ) black hole is a negative cosmological constant black hole in \(d=3\) spacetime dimensions~\cite{Banados:1992wn,Banados:1992gq}. It shares similar thermodynamic properties with other black holes such as those in the AdS. The BTZ black hole entropy is described analogous to the \(d=4\) case with the surface area replaced by the black hole circumference. Similarly, the thermodynamic volume \(V\) scales with the specific volume \(v\) as $ v \sim \sqrt{V} $ as opposed to \(v \sim V^{1/3}\) for the four dimensional black holes. We now describe Hamiltonian flows in the thermodynamic phase space for the BTZ black hole. Consider a contact Hamiltonian function \(h = TS - 2PV\). Then the contact Hamiltonian vector field \(X_h\) takes the form:
\begin{equation}\label{vectorfieldBTZ}
  X_h =  T\frac{\partial}{\partial T} + 2V\frac{\partial}{\partial V} + S\frac{\partial}{\partial S} - 2P\frac{\partial}{\partial P} \, ,
\end{equation}
so that the flow of \(X_h\) is given by:
\begin{equation}\label{flowBTZ}
 \dot{T}=-T, \hspace{3mm} \dot{V}=2V, \hspace{3mm} \dot{S}=S, \hspace{3mm} \dot{P}=-2P \, .
\end{equation}
In particular, the evolution of the \((P,V,T)\) variables is obtained from eqns (\ref{flowBTZ}):
\begin{equation}\label{flowBTZsol}
 T(\tau)=T_0e^{-\tau}, \hspace{3mm} V(\tau)=V_0e^{2\tau}, \hspace{3mm} P(\tau)=P_0e^{-2\tau}; \hspace{5mm} \tau \in \R \, .
\end{equation}
Since specific volume goes as \( v \sim \sqrt{V} \), we may as well write \(v(\tau)=v_0e^\tau\). Then one can easily check that the flows satisfy the well known equation of state for the charged BTZ black hole:
\begin{equation}\label{BTZeqnofstate}
  P = \frac{T}{v} + \frac{Q^2}{2\pi v^2} \, ,
\end{equation}
where in the fixed charge ensemble \(Q=Q_0\).\\

\noindent
From the Smarr relation [eqn (\ref{Smarr})] in \(d=3\), we see that: $h=TS -2PV=0$ and hence the Legendre submanifold representing the BTZ black hole is invariant to the flow of \(X_h\) [eqn(\ref{vectorfieldBTZ})]. Therefore the contact Hamiltonian we considered describes the thermodynamics of the black hole in the thermodynamics phase space. We should remark that this Hamiltonian description for the BTZ black hole is even valid for black holes with zero electric charge, i.e. \(Q=0\). In that case, the equation of state mimics the ideal gas equation.

\subsection{Mapping: Schwarzschild to charged black holes in AdS}

We now consider deformations of the equation of state [eqn (\ref{Schwarzchildeqn})] of the Schwarzschild black hole, that give us corresponding relations for charged AdS black holes. It is important to note that, analogous to the case of van der Waals gas (subsection 2.1.2), such deformations are induced by a contact vector field \(X_h\) generated by some function $h$. The generic vector field \(X_h\) is no longer tangent to the Legendre submanifold representing the Schwarzschild black hole and hence the submanifold is not invariant to the flow of \(X_h\). In this case we can no longer consider \(X_h\) to be the generator of the thermodynamic process but rather as the generator of a family of 1-parameter thermodynamic systems or Legendre submanifolds.
Consider for instance the contact Hamiltonian function:
\begin{equation}\label{AdSg}
  h = -CV^{-1/3}; \hspace{3mm} C > 0 \, ,
\end{equation}
where \(C\) is an arbitrary constant independent of \(V\). Then it can be checked that under the flow of the corresponding contact vector field, \(V\) and \(T\) do not evolve and \(P\) evolves as:
\begin{equation}\label{AdSX}
  \dot{P} = \frac{C}{3}V^{-4/3} \, .
\end{equation}
In \(d=4\), the specific volume is \(v = 2(3V/4\pi)^{1/3}\) thus giving us the deformation in pressure as:
\begin{equation}\label{P}
  P = P_0 + \frac{16C\tau}{3v^4}\bigg(\frac{3}{4\pi}\bigg)^{4/3}; \hspace{5mm} \tau \in \R \, .
\end{equation}
Since \(V\) and \(T\) do not evolve under the flow of \(X_h\), we may as well write \(v=v_0\) and \(T=T_0\). Then we get the deformed equations of state from eqn (\ref{Schwarzchildeqn}):
\begin{equation}\label{AdSA}
  \bigg(P + \frac{1}{2\pi v^2} - \frac{16C\tau}{3v^4}\bigg(\frac{3}{4\pi}\bigg)^{4/3}\bigg)v = T \, .
\end{equation}
We now note that the arbitrary constant \(C\) is actually the charge $Q$ of the black hole.  In particular, if we choose
\begin{equation} \label{Cvalue}
C = \frac{3Q^2}{8\pi}\bigg(\frac{4\pi}{3}\bigg)^{4/3} \, ,
\end{equation}
we exactly reproduce the equation of state for charged black holes in AdS:
\begin{equation}\label{AdSeqn}
  P = \frac{T}{v} - \frac{1}{2\pi v^2} + \frac{2Q^2}{\pi v^4}\tau, \hspace{5mm} \tau \in \R \, .
\end{equation}
We have actually obtained a one-parameter family of charged black holes. As discussed in section-(\ref{ebh}), for the case of static black holes under consideration, entropy and volume are not independent and related to each other in $d=4$ as:
\begin{equation}
V = \frac{4}{3 \sqrt{\pi}}\, S^{\frac{3}{2}} \, .
\end{equation}
Thus, an equivalent contact Hamiltonian following from eqn. (\ref{AdSg}) is:
\begin{equation}\label{AdSgs}
  h = -C\,\left(\frac{4}{3 \sqrt{\pi}} \right)^{\frac{1}{3}} \, S^{-1/2}; \hspace{3mm} C > 0 \, .
\end{equation}
Here, $S$ and $P$ do not vary, but the deformation of temperature $T$ and enthalpy $H$ can be obtained to be:
\begin{eqnarray} \label{THflow}
H &=&  H_0 +  C\, \left( \frac{3\sqrt{\pi}}{4}\right)^{1/3} \, S^{-1/2}\tau,  \nonumber \\
T &=&  T_0 -  \frac{C}{2}\, \left( \frac{3\sqrt{\pi}}{4}\right)^{1/3} \, S^{-3/2}\tau.
\end{eqnarray}
where \(\tau \in \R\) and $T_0, H_0$ are the undeformed quantities listed in eqn. (\ref{Schwarzchild}). Using the value of $C$ obtained earlier, from eqn. (\ref{Cvalue}), we see that $T$ and $H$ derived in eqn. (\ref{THflow}) are exactly the ones corresponding to Reissner-Nordstr\"{o}m charged black holes in AdS given in eqns. (\ref{HRN}) and (\ref{TRN}), respectively.
Therefore, the contact Hamiltonian in eqn. (\ref{AdSg}), maps a Schwarzschild black hole to a family of charged black holes in AdS for this particular choice of \(C\).  We should however remark that this method is not restricted to \(d=4\) dimensions. One can in principle work in higher or even lower dimensions and perform similar deformations to obtain thermodynamic relations and equations of state of more complicated black hole systems.  An appropriate deformation parameter of course needs to be chosen in the contact Hamiltonian, which in the present case involved charge $Q$.

\subsection{Deformations of ideal gas equations}
In the high temperature limit, black holes admit a thermodynamic behaviour similar to that of an ideal gas \cite{Johnson:2015ekr}. In \(d\) dimensions, the high temperature equation of state of a black hole is given by:
\begin{equation}\label{gasideal}
  PV^{1/d-1} = \frac{d-2}{4} \,   \left( \frac{\omega_{d-2}}{ (d-1)} \right)^{1/(d-1)}  T \, ,
\end{equation}
where \(\omega_{d-2}\) is given by:
\begin{equation}\label{omega}
 \omega_{d-2} = \frac{2\pi^{(d-1)/2}}{\Gamma((d-1)/2)} \, .
\end{equation}

In terms of the specific volume \(v\) in \(d\)-dimensions, the equation of state can be shown to take the following simple form:
\begin{equation}
  Pv = T \, .
\end{equation}
By introducing suitable contact vector fields in the thermodynamic phase space, we shall map this high temperature limit to Schwarzschild and charged AdS black holes by deforming eqn (\ref{gasideal}) in \(d=4\). Hence, we perform a similar deformation of the \(d=3\) ideal gas limit which corresponds to the BTZ black hole and map it to charged BTZ black holes.
\subsubsection{Mapping: Ideal gas to Schwarzschild black holes}
To map the ideal gas case to the Schwarzschild black hole, we take:
\begin{equation}\label{idealtoschwarzschildg}
  h = -CV^{1/3}; \hspace{5mm} C > 0 \, ,
\end{equation}
as the deforming contact Hamiltonian where \(C\) is a constant independent of volume. It is not hard to show that \(T\) and \(V\) are conserved along the flow of \(X_h\). The deformation of pressure is calculated to be:
\begin{equation}
  \dot{P} = -\frac{CV^{-2/3}}{3} \, .
\end{equation}
The deformation in pressure expressed in terms of the specific volume \(v\) in \(d=4\) is therefore given by:
\begin{equation}\label{Pideal}
  P = P_0 - \frac{C\tau}{3v^2}\bigg(\frac{6}{\pi}\bigg)^{2/3}; \hspace{5mm} \tau \in \R \, ,
\end{equation}
where \(\tau\) is some real parameter as before. The ideal gas equation written in terms of the specific volume then takes the following form under this deformation:
\begin{equation}\label{s}
  \bigg(P + \frac{C\tau}{3v^2}\bigg(\frac{6}{\pi}\bigg)^{2/3}\bigg)v = T \, .
\end{equation}
For the choice of \(C\): \begin{equation}C = \frac{3}{2 \pi}\bigg(\frac{\pi}{6} \bigg)^{2/3} \, ,\end{equation}
we get the equation of state:
\begin{equation}\label{s2}
  P  = \frac{T}{v} - \frac{\tau}{2\pi v^2}; \hspace{5mm} \tau \in \R \, .
\end{equation}
We have thus mapped the high temperature equation of state to a one-parameter family of Schwarzschild black holes.
\subsubsection{Mapping: Ideal gas to charged black holes}
We now map the ideal gas equation to the equation of state of charged black holes in AdS. For that we consider the contact Hamiltonian of the form:
\begin{equation}
  h = -CV^{1/3} - DV^{-1/3}; \hspace{5mm} C,D > 0 \, ,
\end{equation}
with constants \(C\) and \(D\). The dynamics of pressure \(P\) under the flow of the corresponding contact vector field is given by:
\begin{equation}
  \dot{P} = -\frac{1}{3}\bigg(CV^{-2/3} - DV^{-4/3}\bigg) \, .
\end{equation}
In terms of the specific volume \(v\) in \(d=4\):
\begin{equation}\label{P1}
  P = P_0 -\frac{1}{3}\bigg(CV^{-2/3} - DV^{-4/3}\bigg)\tau; \hspace{5mm} \tau \in \R \, .
\end{equation}
We choose \(C\) and \(D\) to be:
\begin{equation} C=\frac{3}{2\pi}\bigg(\frac{\pi}{6}\bigg)^{2/3} \hspace{5mm} D=\frac{6Q^2}{\pi}\bigg(\frac{\pi}{6}\bigg)^{4/3} \, .\end{equation}
It is not difficult to show that \(T\) and \(V\) are conserved along this flow so that from the ideal gas equation we arrive at the deformed equation of state:
\begin{equation}
  P = \frac{T}{v} - \frac{\tau}{2\pi v^2} + \frac{2Q^2}{\pi v^4}\tau, \hspace{5mm} \tau \in \R \, .
\end{equation}
Therefore the \(X_h\) considered in this section maps the ideal gas system equation to a one-parameter family of charged black holes in the AdS.

\subsubsection{Mapping: Ideal gas to charged BTZ black holes}
 We shall now map the ideal gas limit in \(d=3\) to the charged BTZ black hole.
Now consider a contact Hamiltonian of the form: \begin{equation}h = C \ln(V); \hspace{5mm} C > 0 \, . \end{equation}
where, \(C\) is a constant independent of volume. In this case too, it is easy to show that \(T\) and \(V\) are conserved while \(P\) evolves along the flow of \(X_h\) as: \begin{equation}
    \dot{P} = \frac{C}{V} \, ,
                                                    \end{equation}
so that:
\begin{equation}
  P = P_0 + \frac{C}{V}\tau \hspace{5mm} \tau \in \R \, .
\end{equation}
Referring to the ideal gas equation, we get the one-parameter deformed equations of state:
\begin{equation}
   P = \frac{T}{v} + \frac{Q^2}{2\pi v^2}\tau \hspace{5mm} \tau \in \R \, ,
\end{equation} for if we identify \(C\) with\footnote{One needs to fix some constant factors in \(C\) to get the exact result.} \(Q^2\). These are deformed equations of state corresponding to charged BTZ black holes. Therefore we have mapped the ideal gas limit in \(d=3\) corresponding to an uncharged BTZ black hole to a 1-parameter family of charged BTZ black holes.

\section{Hamilton-Jacobi equations and characteristic curves} \label{HJeqn}
Thus far in this paper, we presented a contact geometry framework for studying thermodynamics of black holes in AdS and related systems, owing to the odd dimensionality of the thermodynamic phase space. Thermodynamic processes can be understood from the mechanics point of view as being trajectories on equilibrium surfaces in a given contact manifold, which defines the dynamical system. Such surfaces are the Legendre submanifolds that take a role analogous to Lagrangian submanifolds or simply configuration spaces in classical mechanics. Therefore as it is done in classical mechanics one may as well study the dynamics only in terms of the configuration space variables i.e. setup a thermodynamic Hamilton-Jacobi theory. From the mechanics point of view, Hamilton-Jacobi equations combined with canonical transformations (which correspond to Legendre transformations in the contact geometry set up) form a powerful combination to solve mechanical problems. It is thus interesting to set up HJ equations for thermodynamics leading to further formal developments, including setting the stage for studying quantum aspects of thermodynamics~\cite{Rajeev2}. Preliminary investigation for setting up HJ equations were already performed in~\cite{Rajeev1}. Here, the equilibrium surfaces following from the equations of state ideal gas and van der Waals gases were used to set up a HJ equation. In particular, attempts were made in~\cite{Rajeev1}, to obtain similar equations for the black holes in AdS. Also, the HJ equation could be integrated in simple situations, with the integration constant being the cosmological constant. We extend the results in~\cite{Rajeev1} to set up HJ equation for black holes in AdS, as appropriate pressure and volume terms are now available in the extended thermodynamic set up. \\

\noindent
Let us consider the case of Reissner-Nordstr\"{o}m black holes in AdS in arbitrary dimension, with equation of state  written as~\cite{Dolan:2010ha}:
\begin{eqnarray} \label{eqnstate}
&& P = \frac{d-2}{16\pi} \, \left[ 4 \pi T \left( \frac{\omega_{d-2}}{V (d-1)} \right)^{1/(d-1)}  - (d-3) \left(\frac{\omega_{d-2}}{V (d-1)} \right)^{2/(d-1)} \right. \nonumber \\
&&~~~~~~~~~~~\left. + (d-3) Q^2 \left( \frac{\omega_{d-2}}{V (d-1)} \right)^{2(d-2)/(d-1)} \right] \, .
\end{eqnarray}
Here, $\omega_{d-2}$ is given by eqn (\ref{omega}). Now, using the relations between extensive and intensive quantities,
\begin{equation} \label{BTZTV}
T = \frac{\partial H}{\partial S}, \qquad V = \frac{\partial H}{\partial P} \, ,
\end{equation}
in equation of state, in eqn. (\ref{eqnstate}) one ends up with the following HJ equation:
\begin{eqnarray} \label{HJGen}
&& P = \frac{d-2}{16\pi}\left[4\pi \left(\frac{\partial H}{\partial S} \right) \left( \frac{\omega_{d-2}}{ (d-1)} \right)^{\frac{1}{(d-1)}}  \left(\frac{\partial H}{\partial P} \right)^{-\frac{1}{(d-1)}} - (d-3) \left(\frac{\omega_{d-2}}{ (d-1)} \right)^{\frac{2}{(d-1)}} \left(\frac{\partial H}{\partial P} \right)^{-\frac{2}{(d-1)}}  \right. \nonumber \\
&&~~~~~~~~~~~~~~ \left. + (d-3) Q^2 \left(\frac{\omega_{d-2}}{ (d-1)} \right)^{\frac{2(d-2)}{(d-1)}} \left(\frac{\partial H}{\partial P} \right)^{\frac{2(d-2)}{(d-1)}} \right]\, .
\end{eqnarray}
Enthalpy $H = H (S,P)$ of the black hole is the solution of the above HJ equation. To show that the set up works, we first invoke the high temperature limit, where the equations of state of black holes are known to reduce to ideal gas equations in general~\cite{Johnson:2015ekr}.  Thus, we consider equations of state of the form:
\begin{equation} \label{id}
P \, V^{1/\gamma}  = c \, T \, ,
\end{equation}
where
\begin{equation}\label{c}
c = \frac{d-2}{4} \,   \left( \frac{\omega_{d-2}}{ (d-1)} \right)^{1/(d-1)}  \, , \qquad  \gamma = d-1 \, .
\end{equation}
are constants which depend on the dimensionality of  the black hole under consideration. The relation in eqn (\ref{id}) is, in fact, the ideal gas limit of the equation in eqn (\ref{eqnstate}), giving the familiar isothermal curves in the $P-v$ plane, where $v \sim V^{1/\gamma}$. As the temperature is lowered, the black hole undergoes wide ranging changes in its phase structure, such as, van der Waals and reentrant phase transitions akin to usual fluids, which has been explored extensively in literature~\cite{Kubiznak:2016qmn}. The corresponding HJ equation can be seen to be:
\begin{equation}
P \left( \frac{\partial H}{\partial P}  \right)^{1/\gamma} = c \,  \left( \frac{\partial H}{\partial S}  \right) \, .
\end{equation}
This PDE can be integrated by separation of variables, giving the \(d\)-dimensional Enthalpy as:
\begin{equation} \label{btzh}
H = \alpha S^{\frac{\gamma}{\gamma-1}}  P \, ,
\end{equation}
with \(\alpha\) given by:
\begin{equation}\label{alpha}
  \alpha = \bigg(\frac{4}{d-1} \bigg)^{\frac{d-1}{d-2}}\bigg( \frac{d-1}{\omega_{d-2}} \bigg)^{1/(d-2)} \, .
\end{equation}
The temperature and volume follow from earlier relations in eqn. (\ref{BTZTV}) to be:
\begin{eqnarray} \label{btztp}
T &=& \frac{\gamma}{\gamma-1}\alpha\, S^{\frac{1}{\gamma-1}}  P \, \, , \nonumber \\
V & = & \alpha S^{\frac{\gamma}{\gamma-1}}  \, .
\end{eqnarray}
The above analysis works in any dimension. For instance, for the case of BTZ black hole in three dimensions, it is known that the equation of state is exactly $P \, V^{1/\gamma}  =  \frac{\sqrt{\pi}}{4} \, T \,$, with $\gamma = 2,\omega_1 = 2 \pi$. It can be checked that thermodynamic relations in eqns. (\ref{btzh}) and (\ref{btztp}) match the existing results in this case~\cite{Dolan:2010ha}.  For more general cases, away from the ideal gas limit, it is in general difficult to integrate the HJ equation completely. Consider for instance the HJ equation for Schwarzschild black hole $(Q=0)$ in $d=4$, as seen from eqn. (\ref{HJGen}) to be:
\begin{equation}
\frac{\partial H}{\partial S} = 2\, P\, \left( \frac{3}{4\pi}\frac{\partial H}{\partial P}\right)^{1/3} + \frac{1}{4\pi}\left( \frac{3}{4\pi}\frac{\partial H}{\partial P}\right)^{-1/3} \, .
\end{equation}
Although, the PDE is not readily integrable, the enthalpy in eqn. (\ref{HSCH}) can be seen to be a solution of the above HJ equation. The validity of HJ equations for black hole thermodynamics reduces framework to \(n\) dimensional configuration spaces which are generated by the thermodynamic potential which in this case is the enthalpy rather than the \((2n+1)\) dimensional thermodynamic phase space. As we see now in the following subsection, characteristic curves can be used to model various thermodynamic processes of black holes.

\subsection{Characteristic curves and thermodynamic processes for black holes}\label{TProcesses}
We now describe the construction of thermodynamic processes and cycles for black holes in the contact geometry framework. Different processes are described by different contact Hamiltonians in the sense that a Hamiltonian function, \(h:\mathcal{M} \rightarrow \R\) vanishing on an equilibrium submanifold \(L\) is treated as the generator of a particular thermodynamic process. The flow of the corresponding contact Hamiltonian vector field shall be tangent to the equilibrium submanifold and that the submanifold is then invariant to the flow. Then the integral curves of the contact vector field shall describe a particular thermodynamic transformation of the system and are alternatively called the characteristic curves. In order to construct a cyclic process one considers patching different processes together so as to describe the cycle. We shall first describe the construction of a few thermodynamic processes by suitable contact Hamiltonians. Hence, we shall discuss the Carnot cycle for the black holes in this setting, with the study of thermodynamic processes in the ideal gas limit.\\

\noindent
Let us see how thermodynamic processes are generated by Hamiltonian functions for the simplest case of high temperature limit of black holes, i.e., the ideal gas limit. As a simple example to illustrate the method, we pick
the uncharged BTZ black hole that corresponds to the ideal gas limit for \(d=3\). However, the following discussion can be extended to more general black hole systems. The thermodynamic equation of state for black hole is:
\begin{equation}P\sqrt{V}=\frac{\sqrt{\pi}T}{4} \, ,\end{equation}
with enthalpy given by:
\begin{equation}H=\frac{4}{\pi}S^2P \, .\end{equation} We now describe as examples, isothermal and isobaric processes for the BTZ black hole generated by contact Hamiltonians.

\subsubsection{Isothermal process}
Consider the contact Hamiltonian of the form:
\begin{equation} \label{ch}
h=T \sqrt{\pi V}/2 - 2PV \, .
\end{equation}
it is immediately clear that:
\begin{equation} \dot{T}=-\frac{\partial h}{\partial S}=0 \, . \end{equation}
Hence, the Hamiltonian describes an isothermal process for the black hole. The corresponding flow of the pressure and volume variables are:
\begin{equation}\dot{P}=\frac{T}{4}\sqrt{\frac{\pi}{V}}-2P, \hspace{3mm} \dot{V}=2V \, , \end{equation}
which means that the integral curves for \((P,V,T)\) are given as:
\begin{equation} \label{intc}
P(\tau)=P_0e^{-\tau}, \hspace{3mm} V(\tau)= e^{2\tau}, \hspace{3mm} T(\tau) = T_0; \hspace{5mm} \tau \in \R \, .
\end{equation}
Here \(P_0 = \sqrt{\pi T_0^2/16 V_0} \).
The integral curves [eqn (\ref{intc})] are seen to satisfy the thermodynamic equation of state of the BTZ black hole. Since for the BTZ black hole, \(S=\sqrt{\pi V}/2\), it is easy to see that \(h=0\) \footnote{As a result of the Smarr relation.} on the equilibrium submanifold which is therefore invariant to the flow of the corresponding contact vector field. Therefore, \(h\) can be interpreted as the generator of an isothermal transformation for the BTZ black hole.

\subsubsection{Isobaric process}
For this one considers the contact Hamiltonian to be of the form: \begin{equation}h=TS - \pi T^2/8P \, .\end{equation} It is then clear that the thermodynamic flow equations are:
\begin{equation}
  \dot{P} = 0, \hspace{3mm} \dot{V} = \frac{\pi T^2}{8 P^2}, \hspace{3mm} \dot{T}=-T \, ,
\end{equation}
which means that the integral curves of the corresponding contact vector field are:
\begin{equation}\label{isobaricbtzevolution}
    P(\tau)=P_0, \hspace{3mm} V(\tau) = V_0e^{2\tau}, \hspace{3mm} T(\tau)=T_0e^{-\tau}; \hspace{5mm} \tau \in \R \, .
\end{equation}
We have used \(V=\pi T^2/16P^2\) given by the equation of state to obtain eqn (\ref{isobaricbtzevolution}) and \(V_0\) is given by \(\sqrt{V_0} = \sqrt{\pi}T_0/4P_0\).
It can be verified that the thermodynamic flows satisfy the equation of state of the BTZ black hole. Also, since from the equation of state for the BTZ black hole, \(4P\sqrt{V}=\sqrt{\pi}T\) this means that \(h=0\) on the equilibrium submanifold. Therefore, this contact Hamiltonian is a generator of an isobaric transformation. These results can be generalized to polytropic gases in arbitrary dimensions.\\

\noindent
One can construct similar contact Hamiltonians to describe other thermodynamic processes (for example isochoric-adiabatic transformations and isothermal transformations) for the BTZ black holes. In principle, this manner of generating different thermodynamic processes using contact Hamiltonians on the thermodynamic phase space still holds for other black holes with more complicated equations of state. Such calculations are however not straightforward to perform analytically. We shall now describe thermodynamic cycles using the ideal gas (high temperature) limit of black holes in arbitrary dimensions.

\subsection{Thermodynamic cycles for AdS black holes}
Having described thermodynamic processes we can now describe thermodynamic cycles such as the Carnot's cycle in the black hole regime. Recall that a 4-stroke cycle such as a Carnot cycle is basically joining four reversible processes together to form a cyclic process in the thermodynamic phase space. This means that in our geometric framework one requires four different contact Hamiltonians for the generating transformations for each "leg" of the cycle which shall lead to the complete description of the cycle.
\begin{figure}[h]
	\begin{center}
		\centering
		\includegraphics[width=5.4in]{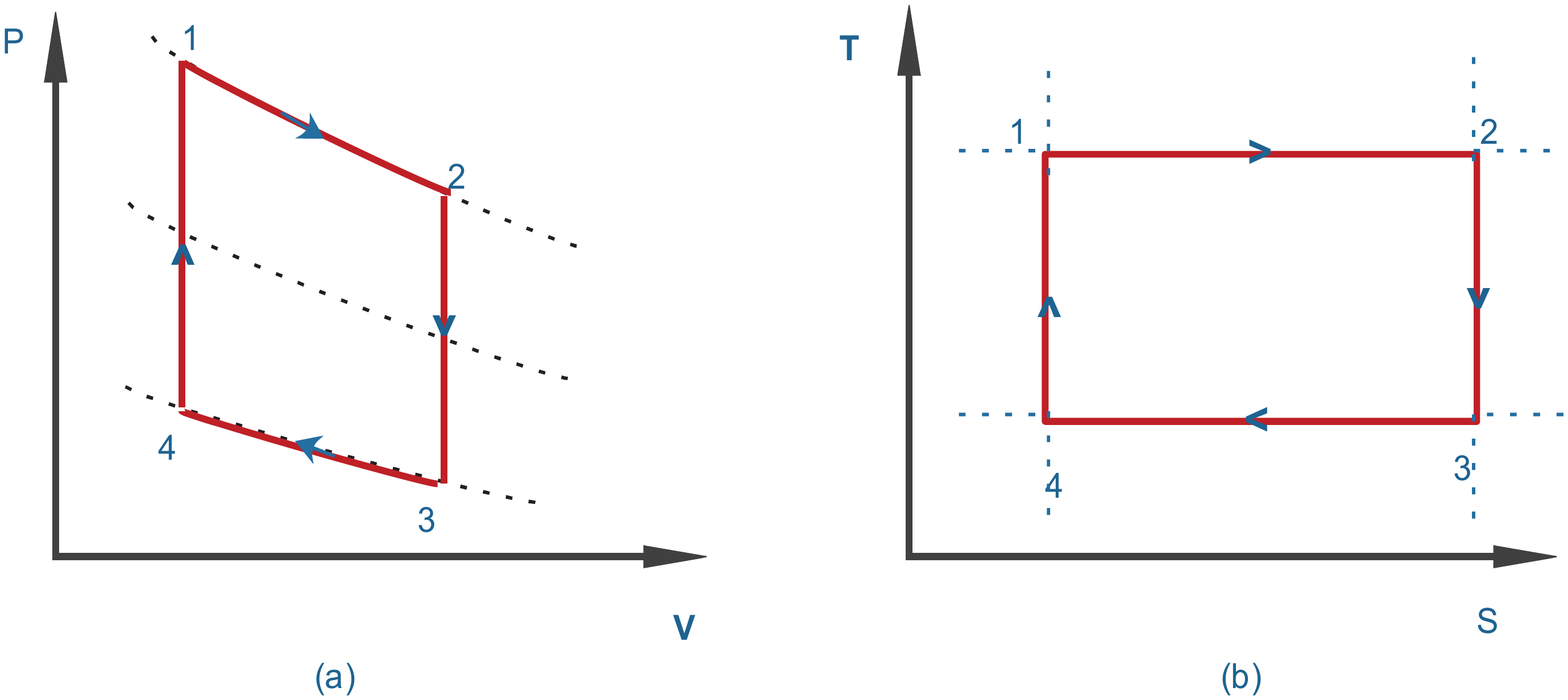}  		
		\caption{Indicator diagrams for a Carnot cycle in the case of black holes: (a) $PV$-plane  (b) $TS$-plane.}   \label{pvtsnew}		
	\end{center}
\end{figure}
We describe the isothermal and adiabatic legs of a Carnot cycle using the high temperature limit of black holes as the working substance, i.e. being described by the equation of state: \(PV^{1/(d-1)}=cT\) with \(c\) being given by eqn (\ref{c}). The corresponding Smarr relation is: \begin{equation}\label{Smarrideal}
                                                    (d-3)H=(d-2)TS - 2PV \, .
                                                  \end{equation}
With some simple manipulations using eqns (\ref{btzh}), (\ref{btztp}) and (\ref{Smarrideal}) it is not difficult to show that in this case: \begin{equation}\label{Smarridealreduced}
(d-2)TS - (d-1)PV=0 \, .
\end{equation}

This constraining relation among thermodynamic variables may be used to construct contact Hamiltonians for generating thermodynamic processes for the Carnot cycle (see Figure \ref{pvtsnew} for indicator diagrams) as we shall now show.

\subsubsection{Isothermal process}
To generate an isothermal transformation, we consider the contact Hamiltonian function:
\begin{equation}\label{idealisoth}
  h =  A_dTV^\frac{d-2}{d-1} - (d-1)PV \, ,
\end{equation}
where $$ A_d = \frac{d-2}{4}(d-1)^\frac{d-2}{d-1}\omega_{d-2}^{\frac{1}{d-1}} \, .$$ is a constant whose value depends on the number of spacetime dimensions one works with. Now it immediately follows that:
\begin{equation}\dot{T} = -\frac{\partial h}{\partial S} = 0 \, .\end{equation}
It can be explicitly verified that the integral curves corresponding to the evolution of the \((P,V,T)\) variables satisfy the thermodynamic equation of state [eqn (\ref{id})]. Moreover, a direct calculation shows that the contact Hamiltonian vanishes\footnote{This follows from the fact that in \(d\)-dimensions the entropy and volume of a static black hole are related as: $$ S = \frac{\omega_{d-2}^\frac{1}{d-1}}{4}[(d-1)V]^\frac{d-2}{d-1} \, .$$ In the \(d=3\) case, all the calculations correspond to those for the BTZ black holes previously described.}, i.e. \(h=0\). Therefore the Legendre submanifold representing the system is invariant to the flow of \(X_h\). Therefore, the contact Hamiltonian in eqn (\ref{idealisoth}) is associated with an isothermal transformation in the high temperature limit of black holes. We shall now describe an adiabatic transformation for the system.

\subsubsection{Adiabatic process}
We now consider constructing an adiabatic transformation for the high temperature black hole. For this we consider the contact Hamiltonian:
\begin{equation}\label{idealadia}
  h = (d-2)c'PV^{1/(d-1)}S - (d-1)PV \, ,
\end{equation}
where, \(c'=1/c\) with \(c\) given by eqn (\ref{c}). One immediately obtains:
\begin{equation} \dot{S} = \frac{\partial h}{\partial T} = 0 \, . \end{equation}
Moreover, the flow of volume is given by:
\begin{equation} \dot{V} = - \frac{\partial h}{\partial P} = (d-1)V - (d-2)c'V^{1/(d-1)}S \, .\end{equation}
A straightforward calculation reveals that\footnote{This follows by substituting for $c'=1/c$ from eqn (\ref{c}) and for \(S\) as a function of \(V\).} \(\dot{V}=0\). Invoking the ideal gas equation [eqn (\ref{id})] into eqn (\ref{idealadia}) it follows that \(h=0\) and hence the Legendre submanifold representing the system is invariant to the flow of \(X_h\). One can verify by explicit calculation that the integral curves satisfy the equation of state. Therefore this contact Hamiltonian [eqn (\ref{idealadia})] is associated with an adiabatic-isochoric transformation. This is natural to expect since for static black holes adiabats are the same as isochores (see Figure \ref{pvtsnew}(a)). For the \(d=3\) case we get an adiabatic-isochoric transformation for the BTZ black hole.\\

\noindent
Working in this manner, one can develop a Hamiltonian formalism for all the four legs of the Carnot cycle. This setup can be extended to describe arbitrary thermodynamic cycles in the thermodynamic phase space. It is worth noting that in the case of static black holes in AdS considered here, even a rectangular cycle in the PV-plane can be chosen for a Carnot process, without loss of generality (see~\cite{Johnson:2014yja} for details). The geometric description of cycles in the thermodynamic phase space for black holes has not been explored yet in the literature to the best of our knowledge.

\section{Remarks} \label{remarks}
In this paper we studied the extended phase space thermodynamics of black holes in AdS using the framework of contact geometry. The contact equations, related vector fields, and characteristic curves were obtained for several systems, such as, Schwarzschild, charged BTZ and Reissner-Nordstr\"{o}m black holes in AdS spacetime, reproducing the well known thermodynamic relations. Hamilton-Jacobi equations of mechanics were then set up for these black holes and their thermodynamic equations of state in the high temperature limit, were shown to follow from the first order differential equations. Once the results are known from high temperature limit, the equations of state of black holes in general, can be obtained from the ideal gas limit via flows of contact vector fields, giving a nice mechanism to generate equations of state of black holes in AdS. Furthermore, we constructed explicit maps of thermodynamic variables as well as the equations of state of black holes in AdS through the deformations introduced by well motivated contact Hamiltonians. We explicitly showed that the thermodynamic relations, including the equations of state of Reissner-Nordstr\"{o}m black holes in four dimensions, can be obtained from those of the Schwarzschild black hole by choosing a contact Hamiltonian with charge as the deformation parameter. The mapping is exact and holds in arbitrary dimensions, as well as for more general black holes. We also showed that the thermodynamic processes of black holes in AdS can be modeled by characteristic curves of a suitable choice of contact Hamiltonian. In this way, isothermal and isobaric processes were explicitly constructed, together with examples involving thermodynamic cycles. The set up is quite general and holds for general black holes in AdS. It would be nice to see whether the maps of equations of state and thermodynamic quantities discussed here, offer deeper insights in understanding further features, such as, mapping of phase transitions and critical exponents in various black hole systems. This should provide an interesting application of the formalism developed here.

\noindent
The analysis in this paper is also valid for non-static i.e. general charged, and rotating black holes in AdS. If the black hole is rotating, one requires additional thermodynamic variables: the angular momentum \(J\) and the angular velocity \(\Omega\). This pair is conjugate to each other and satisfy the Lagrange bracket relationship \(\{J,\Omega\}=1\). The first law of black hole thermodynamics for rotating charged black holes (the Kerr-Newmann black holes) takes the form:
\begin{equation} \label{firstL}
d M = dH = T d S+\Omega d J +\Phi d Q + V dP.
\end{equation}
The corresponding Smarr relation is modified to:$$ (d-3)M=(d-2)TS + (d-2)\Omega J + (d-3)Q\Phi - 2PV \, .$$
This may be used to construct contact Hamiltonian functions for rotating black holes so that the contact flow satisfies the well known thermodynamic equations. For example consider a rotating black hole in \(d=4\). For simplicity we assume that the electric charge is zero (Kerr black hole), i.e. \(Q=0\). We may use the Smarr relation for the rotating case to construct a Hamiltonian function that vanishes on the equilibrium submanifold. It is easy to see that if we take \(h=2TS - 2PV + 2\Omega J - H\) then \(h=0\) and hence the equilibrium submanifold is invariant to the flow of the corresponding \(X_h\). A short calculation reveals that the evolution of the angular momentum is $J(\tau)=J_0e^{2\tau}$ with $\tau \in \R$. It is not hard to check that the well known equation of state (see for instance \cite{Gunasekaran:2012dq}):
$$ P = \frac{T}{v} - \frac{1}{2\pi v^2} + \frac{48J^2}{\pi v^6} \, .$$ is satisfied under the evolution of the thermodynamic variables. It would be interesting to construct contact Hamiltonians for this system and study the deformations induced by it. In this case though, entropy and volume are not independent and it would be nice to obtain the characteristic curves, as would be the case, to explore more general black hole systems in theories with higher derivative terms.\\

\noindent
There are several other avenues to explore for future work. There have been several developments in the study of dissipative systems through contact geometry framework~\cite{CM2}, including geometrical HJ equations, with inputs from canonical transformations. As extensions of the geometric framework studied in this paper, it is interesting to investigate mechanical analogues for black hole systems, such as through a Lagrangian and action formulation.  Some results on Hamiltonian approach to thermodynamic systems were obtained in~\cite{Baldiotti:2016lcf,Baldiotti:2017ywq}, but the analysis was through a different route of symplectic geometry with additional degrees of freedom added by hand, and where, the Lagrangian was a total derivative and non-dynamical. The cosmological constant in these scenarios was shown to be an integration constant~\cite{Baldiotti:2017ywq}, which was also observed in the earlier works~\cite{Rajeev1} though Hamilton-Jacobi approach. In this paper though, we treated the cosmological constant as a dynamical variable, resulting in a $PdV$ term in the first law of black hole mechanics and the existence of an equation of state.
In this approach, in the deformations induced by the contact Hamiltonian on thermodynamic quantities of the black holes, it is the charge of the black hole which is chosen as the constant parameter, eventually leading to maps of various thermodynamic quantities. \\

\noindent
Let us also point out that, although in this work, we restricted ourselves to providing a contact geometry framework to describe black holes in theories with a dynamical and negative cosmological constant, the formalism is very general and readily applicable to situations where the cosmological constant is negative (but non-dynamical), positive or zero. The advantage of studying contact geometry framework in the context of black holes in theories with a negative and dynamical cosmological constant is the availability of a nice equation of state (such as the one in eqn. (\ref{eos1})), which allows for a complete analogy with van der Waals system. Barring this analogy, rest of the computations worked out in this paper, can straightforwardly be generalized to other spacetimes. This is because, as long as a first law (such as the one in eqn. (\ref{firstL})) is available, one can always construct a contact Hamiltonian and study the flows of the resulting vector field, irrespective of whether a $PdV$ term is present in eqn. (\ref{firstL}) or not\footnote{If $PdV$ terms are not present, then, other terms on the R.H.S. of eqn. (\ref{firstL})) play the role of work terms, as is common in traditional black hole thermodynamics.}. It would thus be interesting to generalize these results to study thermodynamics of black holes in other spacetimes.\\

\noindent
Another important issue that requires attention is  the uncertainty relations and quantum treatment of the thermodynamic phase space of black holes. Such relations for ordinary thermodynamics were studied for example in~\cite{Rajeev2,Herczeg:2017xxy}, where uncertainty relations between conjugate pair of thermodynamic variables, such as pressure $P$ and volume $V$ were presented. Similar relations can be proposed in the present case for black holes in AdS too, where the Lagrange bracket structure was collected in eqns (\ref{Lagrange}). The Lagrange brackets are expected to be important while discussing deformation quantization of thermodynamic phase space for black holes, along the lines of~\cite{Rajeev2}.  Another interesting topic to explore is the concept of a heat engine, either in ordinary thermodynamics or in the case of black holes~\cite{Johnson:2014yja}. With the knowledge of thermodynamic processes studied in this paper from geometric point of view, it should be a nice exercise is to construct thermodynamic cycles for more realistic gases and explore their properties. \\

\section*{Acknowledgements}
One of us (A.G.) wishes to acknowledge P. Guha for enlightening discussions on thermodynamic geometry. The authors are grateful to  P. Guha, A. Bravetti and the anonymous referee for a careful reading of the manuscript and valuable comments.

\end{document}